\documentclass[nofootinbib,aps,prd,twocolumn,showpacs,preprintnumbers,amsmath,amssymb]{revtex4}
\usepackage{epsfig}

  \newcommand{\tr}{{\rm Tr}}
\def\lsim{\raise0.3ex\hbox{$<$\kern-0.75em\raise-1.1ex\hbox{$\sim$}}}
\def\gsim{\raise0.3ex\hbox{$>$\kern-0.75em\raise-1.1ex\hbox{$\sim$}}}

\begin{document}

\title{Static quark anti-quark interactions at zero and finite temperature QCD.\\
  II. Quark anti-quark internal energy and entropy}

\author{Olaf Kaczmarek} \email{okacz@physik.uni-bielefeld.de}
\affiliation{ Fakult\"{a}t f\"{u}r Physik, Universit\"{a}t Bielefeld, D-33615
  Bielefeld, Germany }

\author{Felix Zantow} \email{zantow@quark.phy.bnl.gov} \affiliation{Physics
  Department, Brookhaven Natl. Laboratory, Upton, New York 11973, USA}

\date{\today} \preprint{BI-TP 2005/09 and BNL-NT-05/21}


\begin{abstract}
  We analyze the change in free energy, internal energy and entropy due to the
  presence of a heavy quark anti-quark pair in a QCD heat bath. The internal
  energies and entropies are introduced as intensive observables calculated
  through thermal derivatives of the QCD partition function containing
  additional static color sources. The quark anti-quark internal energy and, in
  particular, the entropy clearly signal the phase change from quark
  confinement below and deconfinement above the transition and both observables
  are introduced such that they survive the continuum limit. The intermediate
  and large distance behavior of the energies reflect string breaking and color
  screening phenomena. This is used to characterize the energies which are
  needed to dissolve heavy quarkonium states in a thermal medium. Our
  discussion supports recent findings which suggest that parts of the
  quarkonium systems may survive the phase transition and dissolve only at
  higher temperatures.
\end{abstract}

\maketitle

\section{Introduction}
This is the second part of our discussion of thermal modifications of the
strong forces in finite temperature QCD \cite{Kaczmarek:2005ui} (for a detailed
introduction to this subject and further references see
\cite{Kaczmarek:2005ui,Karsch:2005ex}). At finite temperature, $T\neq0$, the
free energy of a static quark anti-quark pair
\cite{McLerran:1981pb,McLerran:1980pk}, separated by distance $r$, is an
important tool to analyze the in-medium modification of the QCD forces. Similar
to the free energies also the internal energies have recently been introduced
\cite{Kaczmarek:2002mc,Zantow:2003ui} and are expected to play an important
role in the discussion of quarkonia binding properties
\cite{Matsui:1986dk,Brown:2004qi,Wong:2004kn,Shuryak:2004tx,Park:2005nv,
Brambilla:2004wf,Digal:2001iu,Digal:2001ue,Wong:2001kn,Wong:2001uu,Shuryak:2003ty}.
Moreover, the structure of these energies at large distances and high
temperatures is important for our understanding of the bulk properties of the
QCD plasma phase, {\em e.g.} the screening property of the quark gluon plasma
\cite{Kaczmarek:1999mm,Kaczmarek:2004gv}, the equation of state
\cite{Beinlich:1997ia,Karsch:2000ps}. They also provide important input to the
construction of effective models based on properties of the Polyakov loop
\cite{Pisarski:2002ji,Kaczmarek:2002mc,Kaczmarek:2005ui,Kaczmarek:2003ph,Dumitru:2003hp,Dumitru:2004gd}.
Up to quite recently
\cite{Karsch:2000ps,Kaczmarek:2005ui,Kaczmarek:2003ph,Petreczky:2004pz,Kaczmarek:2005uw,Kaczmarek:2005uv}
most of these discussions concerned the quark anti-quark free energies in
quenched QCD. Several qualitative differences, however, are to be expected when
changing from free to internal energies and/or when taking into account the
influence of dynamical fermions. The difference between free and internal
energy arises from non-trivial entropy contributions
\cite{Kaczmarek:2002mc,Zantow:2003ui}. Moreover, in QCD with light dynamic
quarks the large distance behavior of the strong interaction will show a
qualitative different behavior due to the possibility of string breaking.

Properties of the finite temperature quark anti-quark free energies and the
heavy quark potential at $T=0$, $V(r)$, have been discussed for $2$-flavor QCD
in
Refs.~\cite{Karsch:2000ps,Kaczmarek:2005ui,Kaczmarek:2005uw,Kaczmarek:2005uv,Kaczmarek:2003ph}.
Recently some results have also been reported for $3$-flavor QCD
\cite{Petreczky:2004pz}. Here we will continue our analysis of the fundamental
forces of QCD at finite temperature. We analyze the partition function of
$2$-flavor QCD in the presence of heavy quarks and extract the quark anti-quark
internal energies and entropies. This paper is organized as follows: In section
\ref{secen1} we introduce the change in internal energy and entropy due to the
presence of static quarks and anti-quarks in a thermal heat bath. We discuss
their temperature dependence at large distances, in particular, their behavior
in the vicinity of the transition, in section \ref{secen12}. We finally discuss
the qualitative and quantitative differences between free and internal energies
in section \ref{sec3} and discuss their binding properties with respect to
quarkonium binding in the vicinity of the transition. Section~\ref{seccon}
contains our conclusions. Details on our simulation parameters, the lattice
actions used in our calculations as well as details on the analysis of the
quark anti-quark free energies are given in Ref.~\cite{Kaczmarek:2005ui}.

\section{Partition function in the presence of heavy quarks}\label{secen1}
\subsection{Free energy, internal energy and entropy}
As we are interested in the lattice formulation of QCD at finite temperature in
thermal equilibrium, we consider the (Euclidean) path integral, {\em i.e.} we
consider the partition function of the QCD heat bath,
\begin{eqnarray}
{\cal Z}(T,V)&\equiv&\int dAd\Psi d\bar{\Psi} e^{-S[A,\Psi,\bar{\Psi}]}\;=\;e^{-F(T,V)/T},\label{part}
\end{eqnarray}
where $T$ ($V$) denotes the temperature (volume) and $S[A,\Psi,\bar{\Psi}]$ the
QCD action. We also investigate the corresponding system containing additional
heavy quarks \cite{McLerran:1981pb}, {\em i.e.}
\begin{eqnarray}
{\cal Z}_{\cal O}(r,T,V)&\equiv&\int dAd\Psi d\bar{\Psi}\; {\cal
  O}^{(c)}_r[W,W^\dagger]\; e^{-S[A,\Psi,\bar{\Psi}]}\nonumber\\
&=&e^{-\tilde{F}_{{\cal O}}(r,T,V)/T}\;,\label{partqq}
\end{eqnarray} 
where ${\cal O}^{(c)}_r[W,W^\dagger]$ denotes an operator which introduces
static color sources representing quarks and anti-quarks separated by distances
$r\equiv\{r_i\}$ in a specific color representation $c$. A static color source
appearing in ${\cal O}_r^{(c)}[W,W^\dagger]$ located at ${\bf x}$ is described
by the thermal Wilson line,
\begin{eqnarray}
W({\bf x})&=&{\cal T}\exp\left(i\int_0^{1/T}dx_0\;{\bf \lambda} \cdot {\bf A}_0(x_0,{\bf x}) \right)\;.\label{W}
\end{eqnarray}
The expectation value of ${\cal O}^{(c)}_r[W,W^\dagger]$, {\em i.e.} the
$n$-point Polyakov loop correlation function $\langle {\cal
  O}^{(c)}_r[W,W^\dagger]\rangle$, is related to the change in free energy,
$F_{\cal O}(r,T)\equiv \tilde{F}(r,T,V)-F(T,V)$, due to the presence of static
quark anti-quark sources in the heat bath,
\begin{eqnarray}
F_{\cal O}(r,T)&=&-T\ln\langle {\cal O}^{(c)}_r[W,W^\dagger]\rangle+TC_{\cal O}\label{here}\\
&=&-T\left(\ln {\cal Z}_{\cal O}(r,T,V)-\ln{\cal Z}(T,V)\right)+TC_{\cal O}\;,\nonumber
\end{eqnarray}
where $C_{\cal O}$ can be fixed through renormalization \cite{Note1}. For
instance, in the case of $2$-point Polyakov loop correlation functions the
singlet ($1$), averaged ($\bar q q$) and octet ($8$) color representations of
the operator ${\cal O}^{(c)}_r[W,W^\dagger]$ can be specified as
\cite{McLerran:1981pb,Nadkarni:1986as,Philipsen:2002az}
\begin{eqnarray}
{\cal O}^{(1)}_r[W,W^\dagger]&=&\frac{1}{3}\tr\;W(0)\;W^\dagger(|r|)\;,\label{sin}\\
{\cal O}^{(\bar q q)}_r[W,W^\dagger]&=&\frac{1}{9}\tr\;W(0)\;\tr\;W^\dagger(|r|)\;,\label{ave}\\
{\cal O}^{(8)}_r[W,W^\dagger]&=&\frac{1}{8}\tr\;W(0)\;\tr\;W^\dagger(|r|)\nonumber\\
&&-\frac{1}{24}\tr\;W(0)\;W^\dagger(|r|)\;.\label{oct}\end{eqnarray}
The color singlet, averaged and octet quark anti-quark free energies, {\em
  i.e.} $F_1(r,T)$, $F_{\bar q q }(r,T)$ and $F_8(r,T)$, respectively, have
already been discussed extensively in quenched and full QCD
\cite{Attig:1988ey,Kaczmarek:2002mc,Kaczmarek:2004gv,Kaczmarek:2005ui,
Kaczmarek:2003ph,Kaczmarek:2003dp,Petreczky:2004pz,Kaczmarek:2005uw,Kaczmarek:2005uv,Nakamura:2004wr}.

For the purpose of discussing internal energies ($U_{\cal O}(r,T)$) and
entropies ($S_{\cal O}(r,T)$), we follow the conceptual approach suggested in
Refs.~\cite{Kaczmarek:2002mc,Zantow:2003ui} and consider thermal derivatives of
the QCD partition functions introduced above, {\em i.e.}
\begin{eqnarray}
U_{\cal O}(r,T)&=&-T^2\frac{\partial F_{\cal O}(r,T)/T}{\partial T}\;,\label{Ucalc}\end{eqnarray}
which leads to 
\begin{widetext}
\begin{eqnarray}
U_{\cal O}(r,T)&=&-T^2\left(\frac{1}{\langle {\cal O}^{(c)}_r[W,W^\dagger]\rangle}
\langle {\cal O}^{(c)}_r[W,W^\dagger]\;\frac{\partial 
S[A,\Psi,\bar{\Psi}]}{\partial T}\rangle \;-\;\langle \frac{\partial S[A,\Psi,\bar{\Psi}]}{\partial T}\rangle\right)\nonumber\\
&\equiv&\tilde{U}_{\cal O}(r,T,V)-U(T,V)\;.\label{Ulat}
\end{eqnarray}
\end{widetext}
Note here that the derivative of the operator ${\cal O}_{r}^{(c)}$ with respect
to temperature, $\partial {\cal O}^{(c)}_r/\partial T$, vanishes due to
(\ref{W}).  A similar relation can be derived for the entropies starting from
\begin{eqnarray}
S_{\cal O}(r,T)&=&-\frac{\partial F_{\cal O}(r,T)}{\partial T}\nonumber\\
&\equiv&\tilde{S}_{\cal O}(r,T,V)-S(T,V)\;-\;\frac{\partial TC_{\cal O}}{\partial T}\;,\label{Scalc}
\end{eqnarray}
{\em i.e.} the observable $TS_{\cal O}(r,T)$ could be calculated from the
difference, $TS_{\cal O}(r,T)=U_{\cal O}(r,T)-F_{\cal O}(r,T)$, with $U_{\cal
  O}(r,T)$ and $F_{\cal O}(r,T)$ given in (\ref{Ulat}) and (\ref{here}). We
have also specified constant contributions, $C_{\cal O}$, which result from
divergent contributions to the free energies and will control the internal
energies and entropies at large quark anti-quark separations. Once the free
energies are fixed through renormalization also the constant contributions to
the internal energies and entropies are properly determined.

We note that Eq.~(\ref{Ulat}), and similar (\ref{Scalc}), open the possibility
for a direct calculations of $U_{\cal O}(r,T)$ and $S_{\cal O}(r,T)$ and define
properly the quantities we aim to discuss here, {\em i.e.} the change in
internal energies and entropies due to the presence of static quarks and
anti-quarks in the QCD heat bath. Quite similar to the change in free energies,
$F_{\cal O}(r,T)=\tilde{F}_{\cal O}(r,T)-F(T)$, also the changes in internal
energies and entropies, $U_{\cal O}(r,T)=\tilde{U}_{\cal O}(r,T,V)-U(T,V)$ and
$S_{\cal O}(r,T)=\tilde{S}_{\cal O}(r,T,V)-S(T,V)$, are expected to behave like
intensive observables and, in particular, will show no volume dependence in the
thermodynamic limit. It, however, can no longer be assumed that the
$r$-dependence of the quark anti-quark free energies (\ref{here}) are given by
the $r$-dependences of the internal energies (\ref{Ulat}) alone, {\em i.e.} we
expect
\begin{eqnarray}
F_{\cal O}(r,T)&=&U_{\cal O}(r,T)-TS_{\cal O}(r,T)\;.\label{nontriv}
\end{eqnarray}
This indicates a quite complicated relation between free energies, internal
energies and entropies and stresses the important role of the entropy
contribution which is still present in free energies. In the limit of vanishing
temperature, $T\to0$, however, $TS_{\cal O}(r,T\to0)$ will vanish and
\begin{eqnarray}
F_{\cal O}(r,T\equiv0)\;=\;U_{\cal O}(r,T\equiv0)\;\equiv\;V_{\cal O}(r)\;,
\end{eqnarray}
{\em i.e.} $F_{\cal O}(r,T\to0)$ and $U_{\cal O}(r,T\to0)$ could be
used\footnote{We assume here the existence of an energy, $V_{\cal O}(r)$, at
  $T=0$ which corresponds to the expectation value of $\lim_{T\to0}\langle{\cal
    O}_r^{(c)}\rangle$.} to define the corresponding energies at $T=0$.

Unfortunately, the non-perturbative formulation of energies given in
(\ref{Ulat}) is still complicated and, in particular, it is complicated to be
given in a form suitable for lattice simulations
\cite{Karsch:2000ps,Boyd:1996bx,Boyd:1995zg,Montvay:1994cy}. In the
thermodynamic limit, however, the internal energy and entropy described above
can be calculated equally well from Eqs.~(\ref{Ucalc}, \ref{Scalc}). We indeed
have used these relations and performed the derivatives with respect to
temperature based on finite difference approximations using the
non-perturbative free energies at neighboring temperature as input (see Tab.~1
of Ref.~\cite{Kaczmarek:2005ui}). With this method no perturbative
uncertainties get introduced in our calculations and a test of confinement and
deconfinement could be given. In fact, as we use in our calculations the
renormalized free energy as input, both quantities, the internal energies and
entropies, survive the continuum limit. We stress here, however, that we also
provide non-perturbative renormalization prescriptions for the quark anti-quark
internal energies and entropies which are independent from a renormalization of
the quark anti-quark free energies.

\subsection{Theoretic expectations and renormalization}\label{exp}
Following \cite{Kaczmarek:2005ui} we consider mainly the quark anti-quark
internal energies and entropies in the color singlet channel. At large
distances and high temperatures, {\em i.e.} $rT\gg1$, $T$ sufficiently above
$T_c$, as well as at small distances and zero temperature, {\em i.e.}
$r\Lambda_{\text{QCD}}\ll1$, the singlet free energies, $F_1(r,T)$, are indeed
dominated by one gluon exchange \cite{Kaczmarek:2005ui}. Using the perturbative
small distance relation for the singlet free energy
\cite{Brown:1979ya,Nadkarni:1986as,Nadkarni:1986cz,McLerran:1981pb},
\begin{eqnarray}
F_1(r,T)&\simeq&-\frac{4}{3}\frac{\alpha(r)}{r}\quad\;,\;r\Lambda_{\text{QCD}}\ll1\;,\label{alp_rT1}
\end{eqnarray}
standard thermodynamic relations (Eqs. (\ref{Ucalc}) and (\ref{Scalc})) lead to
\begin{eqnarray}
U_1(r\ll1/\Lambda_{\text{QCD}},T)&\simeq&-\frac{4}{3}\frac{\alpha(r)}{r}\;, \label{intren}
\end{eqnarray}
and
\begin{eqnarray}
S_1(r\ll1/\Lambda_{\text{QCD}},T)&\simeq&0\;.\label{entren}
\end{eqnarray}
We thus expect that at small distances the singlet free and internal energies
are controlled to a large extent by energy, {\em i.e.} in the limit of small
distances both will smoothly approach the zero temperature heavy quark
potential, $V(r)$. At larger distances, however, the quark anti-quark free
energies are strongly temperature dependent
\cite{Kaczmarek:2005ui,Kaczmarek:2005uv,Kaczmarek:2005uw} and thus
non-vanishing entropy contributions arise. In this case differences between
free and internal energies are expected to become important and the quark
anti-quark free energy will be to a large extent controlled by $TS_1(r,T)$.

\begin{figure}[tbp]
  \epsfig{file=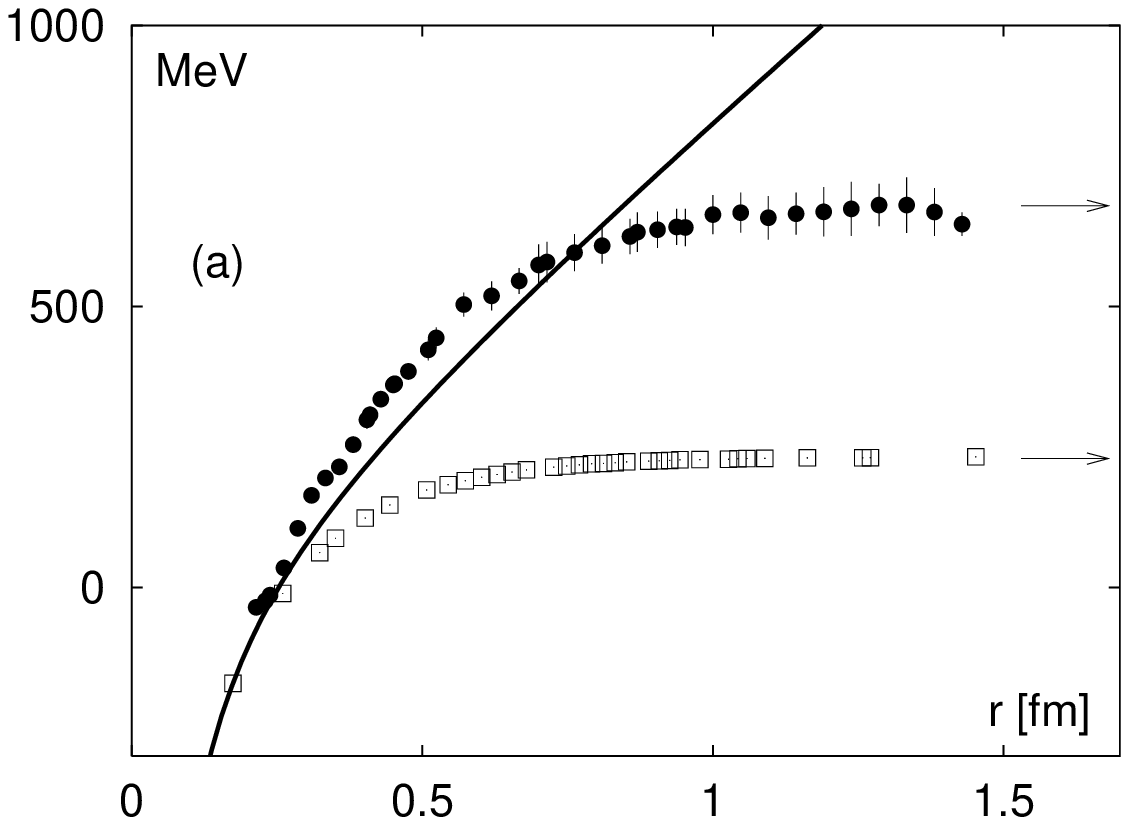,width=9.0cm}
  \epsfig{file=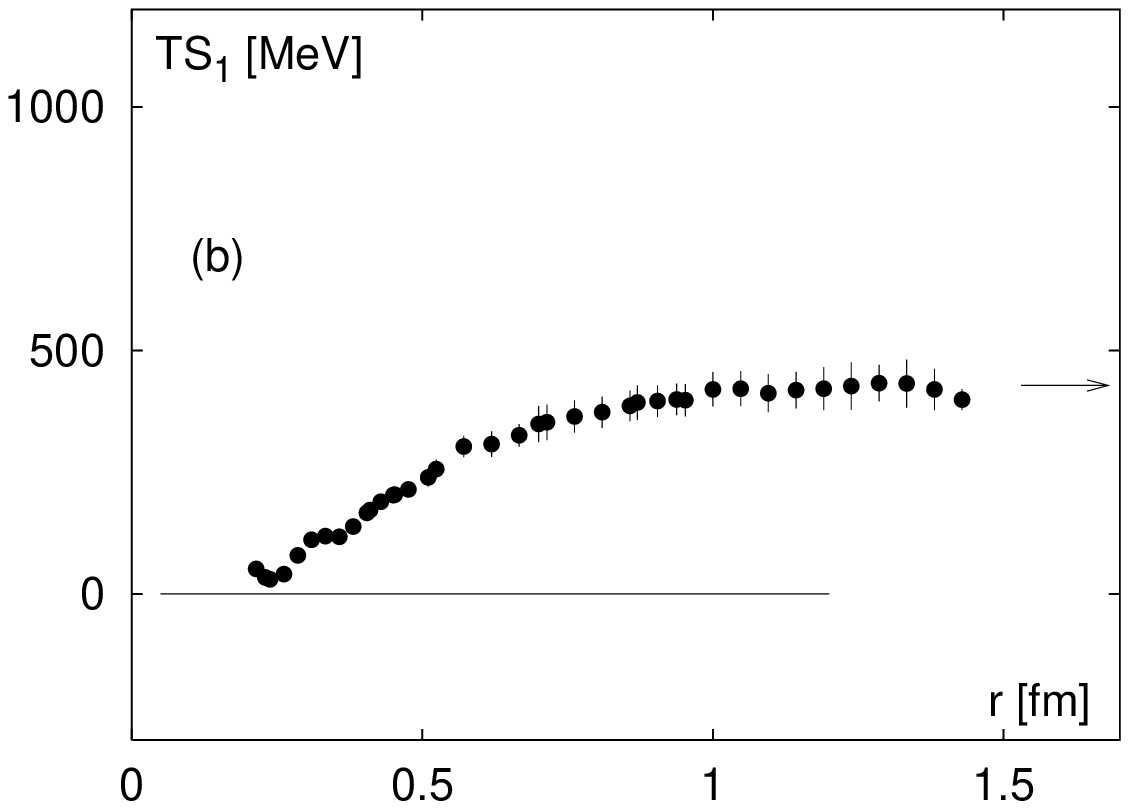,width=9.0cm}
\caption{(a) The singlet internal energy, $U_1(r,T)$ (filled circles),
  calculated from renormalized singlet free energy, $F_1(r,T)$ (open squares),
  at fixed $T\simeq1.3T_c$ calculated in $2$-flavor lattice QCD and compared
  them to $V(r)$ (line) \cite{Kaczmarek:2005ui,Kaczmarek:2005uw}. (b) The
  corresponding color singlet quark anti-quark entropy,
  $TS_1(r,T\simeq1.3T_c)$, as function of distance calculated from renormalized
  free energies. The arrows point in both figures at the temperature dependent
  values of the free and internal energy and entropy at asymptotic large
  distances, {\em i.e.} $F_\infty(T)\equiv \lim_{r\to\infty}F_1(r,T)$,
  $U_\infty(T)\equiv \lim_{r\to\infty}U_1(r,T)$ and $TS_\infty(T)\equiv
  T\lim_{r\to\infty}S_1(r,T)$.  
}
\label{plc_ren}
\end{figure}
To clarify the role of the entropy we compare\footnote{Details on our lattice
  simulations, in particular, on the calculation of $F_1(r,T)$, are given in
  Ref.~\cite{Kaczmarek:2005ui}. Details on the computation of $U_1(r,T)$ and
  $S_1(r,T)$ will be given in Sec.~\ref{secen12} and \ref{sec3}.} in
Fig.~\ref{plc_ren}(a) the short and large distance parts of the singlet free
and internal energies at $T\simeq1.3T_c$ and show in Fig.~\ref{plc_ren}(b) the
corresponding entropy contribution, $TS_1(r,T\simeq1.3T_c)$. We also indicate
in both figures the small distance behavior expected from Eqs.~(\ref{intren})
and (\ref{entren}) as solid lines, {\em i.e.} the line in Fig.~\ref{plc_ren}(a)
indicates the heavy quark potential from
Refs.~\cite{Kaczmarek:2005ui,Kaczmarek:2005uw}, and in Fig.~\ref{plc_ren}(b) it
indicates the zero level. As both, the internal energy and entropy, have been
calculated using renormalized free energies proper renormalization of both
observables is already incorporated by construction. It can clearly be deduced
from Fig.~\ref{plc_ren}(a) that the singlet free and internal energies smoothly
approach $V(r)$ at small distances and the free energy thus indeed is dominated
by the energy contribution. In fact, the entropy contribution shown in
Fig.~\ref{plc_ren}(b) is quite small at small distances and indicates a
vanishing entropy contribution in the limit $r\to0$. Unfortunately we could not
go to smaller distances to clearly demonstrate this behavior. Moreover, at
small distances $U_1(r,T)$ and $TS_1(r,T)$ suffer from lattice artifacts which
result from small distance lattice artifacts in the free energies. At
intermediate and large distances, however, the free and internal energies shown
in Fig.~\ref{plc_ren}(a) deviate from each other and $TS_1(r,T)$ indeed plays
an important role for the behavior of $F_1(r,T)$. At asymptotic large distances
the internal energies and entropies approach temperature dependent constant
values, {\em i.e.} $U_\infty(T)\equiv\lim_{r\to\infty}U_1(r,T)$ and
$S_\infty(T)\equiv\lim_{r\to\infty}S_1(r,T)$ are finite for finite
temperatures. These values are indicated by the arrows in Fig.~\ref{plc_ren}
and, in particular, at finite temperature
\begin{eqnarray}
U_\infty(T)&\gsim& F_\infty(T)\;\label{ulf}
\end{eqnarray}  
is evident.

A similar behavior of $U_\infty(T)$ and $S_\infty(T)$ can be deduced from high
temperature perturbation theory. To be more precise, high temperature
perturbation theory suggests a cubic leading order dependence\footnote{Here and
  in what follows we already have anticipated the running of the coupling with
  the expected dominant scale $T$. Of course, the running of the coupling
  appears only beyond leading order.} of the free energy on the coupling
\cite{Gava:1981qd}, {\em i.e.}
\begin{eqnarray}
F_\infty(T)\simeq-\frac{4}{3} m_D(T)\alpha(T)\;\simeq\;-{\cal O}(g^3T)\;.\label{Finfty}
\end{eqnarray}
Here $\alpha(T)=g^2(T)/4\pi$ and $m_D(T)$ denotes the Debye mass which in
re-summed leading order is given by
\begin{eqnarray}
m_D(T)&=&\left(1+\frac{N_f}{6}\right)^{1/2}\;g(T)T\;.
\end{eqnarray}
We note that this leading order result is gauge invariant. In the following we
use the renormalization group $\beta$-function to evaluate the derivatives of
the coupling, {\em i.e.} for an arbitrary function ${\cal F}\equiv{\cal
  F}(g,T)$ we use
\begin{eqnarray}
T\frac{d{\cal F}(g,T)}{dT}&=&T\frac{\partial{\cal F}(g,T)}{\partial T}+\beta(g)\frac{d{\cal F}(g,T)}{dg}\;,
\end{eqnarray}
where $\beta(g)=-\beta_0g^3+{\cal O}(g^5)$ in perturbation theory.  Assuming
this behavior the internal energy, $U_\infty(T)$, and entropy, $S_\infty(T)$,
are expected to behave like
\begin{eqnarray}
U_\infty(T)&\simeq&4 m_D(T)\alpha(T)\frac{\beta(g)}{g(T)}\;\simeq\;-{\cal O}(Tg^5)\;,\label{Upert}
\end{eqnarray}
and
\begin{eqnarray}
S_\infty(T)&\simeq&+\frac{4}{3}\frac{m_D(T)}{T}\alpha(T)+4\frac{m_D(T)}{T}\alpha(T)\frac{\beta(g)}{g(T)}\nonumber\\
&\simeq&+{\cal O}(g^3)\;.\label{entlimit}
\end{eqnarray}  
The leading contribution to $TS_\infty(T)$ is similar to the free energy in
Eq.~(\ref{Finfty}) and thus at high temperatures and large distances the free
energy is indeed expected to be to large extent dominated by the entropy
contribution, {\em i.e.} at leading order $S_\infty(T)\simeq-F_\infty(T)/T$.
Although the entropy itself will vanish logarithmically in the high temperature
limit, {\em i.e.} $S_\infty(T\to\infty)=0$, the contribution $TS_\infty(T)$
will clearly dominate the differences between free and internal energy at high
temperatures,
\begin{eqnarray}
U_\infty(T)-F_\infty(T)&=&TS_\infty(T)\;\simeq\;+{\cal O}(g^3T)\;. \label{ts}
\end{eqnarray}
Thus, the difference between free and internal energy is expected to increase
continuously with increasing temperature when approaching the perturbative high
temperature regime. Only in the limit of zero temperature, $T\to0$, the
observable $TS_\infty(T)$ will vanish as $S_\infty(T)$ is a dimension less
quantity which due to string breaking stays finite in QCD. Any qualitative
change in the observable $TS_\infty(T)$ as function of temperature between both
limits, {\em i.e.} $T\to0$ and $T\to\infty$, is not quite obvious and, if
present, may signal the phase change from the chiral symmetry broken phase at
low to the deconfinement phase at high temperatures.

We may finally note that although we discuss in the following the internal
energies and entropies calculated from renormalized free energies, it is
conceptually quite satisfying that both observables could equally well be
renormalized by matching their short distance parts to the heavy quark
potential (\ref{intren}) and zero (\ref{entren}), respectively. This is indeed
evident from Figs.~\ref{plc_ren}(a, b). As no additional divergences get
introduced at finite temperature also their large distance properties are
properly fixed in the continuum limit, in particular, also the manifestly gauge
invariant observables $U_\infty(T)$ and $S_\infty(T)$.

\section{The confinement deconfinement transition}\label{secen12}
We begin our discussion of the finite temperature energies and entropies at
asymptotic large distances, $r\to\infty$. Actually, to avoid any fit we again
\cite{Kaczmarek:2005ui} approximated the value of the free energy at infinite
distance, $F_1(r\to\infty,T)$, by the value of the quark anti-quark free energy
calculated at the largest possible separation on the lattice, {\em i.e.}
$F_\infty(T)\equiv F_{\bar qq}(N_\sigma/2,T)$, and calculated separately the
internal energy, $U_\infty(T)$, and entropy, $S_\infty(T)$. Both quantities are
obtained from derivatives of the color averaged free energy, $F_{\bar q
  q}(r,T)$, which is a manifestly gauge invariant observable. Our results for
$U_\infty(T)$ and $S_\infty(T)$ are summarized in Tab.~\ref{tab1}. It is quite
satisfying that the values obtained for $U_\infty(T)$ and $S_\infty(T)$
reproduce the free energy given in Tab.~2 of Ref.~\cite{Kaczmarek:2005ui}, {\em
  i.e.} the quantity $U_\infty-TS_\infty$ matches to the value for
$F_\infty=-2T\ln|\langle L\rangle|$ for all temperatures.

\begin{table}[tbp]
\centering
\setlength{\tabcolsep}{0.7pc}
\begin{tabular}{|l|l|l|}
\hline
$T/T_c$ & $S_\infty$ & $U_\infty/T_c$ \\
\hline
\hline

0.79 &  5.48  (259) & 9.37  (207)\\
0.84 &  6.49  (119) & 10.19 (101)\\
0.88 &  7.78  (168) & 11.30 (148)\\
0.93 &  12.92 (138) & 15.96 (129)\\
0.98 &  16.38 (113) & 19.27 (112)\\
1.01 &  14.83 (172) & 17.71 (173)\\
1.04 &  12.93 (825) & 15.78 (88)\\
1.09 &  5.49  (42) & 7.84  (46)\\
1.13 &  3.95  (41) & 6.14  (46)\\
1.19 &  2.73  (25) & 4.72  (29)\\
1.29 &  1.63  (16) & 3.37  (21)\\
1.43 &  1.25  (11) & 2.85  (16)\\
1.57 &  1.02  (8) & 2.51  (12)\\
1.72 &  0.91  (10) & 2.32  (17)\\
1.89 &  0.87  (7) & 2.26  (14)\\
2.99 &  0.67  (1) & 2.09  (2)\\

\hline
\end{tabular}
\caption{The change in internal energy and entropy due to the presence of a
  quark anti-quark pair at infinite separation in the QCD heat bath versus
  temperature.}
\smallskip
\label{tab1}
\end{table}

In parts of our analysis we are also interested in the flavor and quark mass
dependence of the finite temperature energies and entropies. We thus again
\cite{Kaczmarek:2005ui} compare our results ($N_f=2$) to results from quenched
QCD ($N_f=0$) \cite{Kaczmarek:2002mc,Phd} and also in parts to a recent study
of $3$-flavor QCD \cite{Petreczky:2004pz}. To convert the observables to
physical units we use $T_c=270$ MeV in quenched, $T_c=200$ MeV in $2$-flavor
($m_\pi/m_\rho\simeq0.7$) and $T_c=193$ MeV in $3$-flavor
($m_\pi/m_\rho\simeq0.4$) QCD. It should be obvious, however, that a comparison
of free and internal energies crucially depends on the relative normalization
of the corresponding zero temperature heavy quark potentials used for
renormalization and thus a comparison could be affected by flavor and/or quark
mass dependent (over-all) constant contributions. Here, and in what follows,
the relative normalization of the heavy quark potentials in quenched and full
QCD is such that there is no constant contribution in the Cornell Ansatz for
$V(r)$ at large distances. We also note that any undetermined constant
contribution to the heavy quark potential at zero temperature will add a
non-perturbative over-all constant to the free and internal energies which
would also affect the comparison of these observables with perturbation theory
\cite{Zantow:2003uh}. We stress again, however, that the quark anti-quark
entropy is unaffected by any undetermined finite renormalization of $V(r)$ at
zero temperature, {\em i.e.} $S_\infty(T)$ does not dependent on any flavor
and/or quark mass dependent normalization terms that could contribute to $V(r)$
at $T=0$.

\subsection{The free energy}\label{FE}
\begin{figure}[bp]
  \epsfig{file=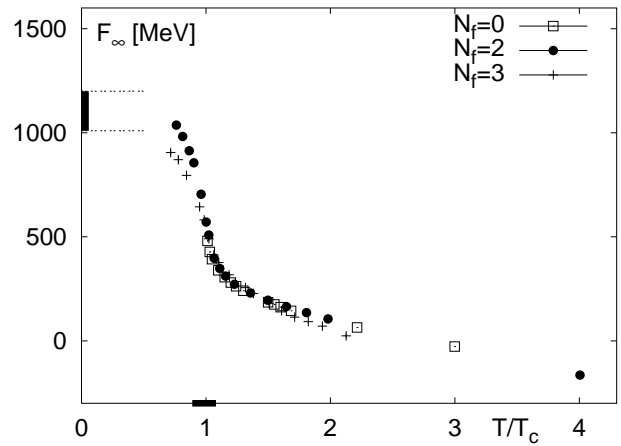,width=9.0cm}
\caption{
  The value of the free energies $F_\infty(T)$ at asymptotic large distances as
  function of temperature in physical units. We compare here our results from
  $2$-flavor QCD ($T_c=200$ MeV, $m_\pi/m_\rho\simeq0.7$) to the results in
  pure gauge theory ($T_c=270$ MeV) \cite{Kaczmarek:2002mc} and $3$-flavor QCD
  ($T_c=193$ MeV, $m_\pi/m_\rho\simeq0.4$) \cite{Petreczky:2004pz}. The dashed
  horizontal lines show the expected energy
  $V(r_{\text{breaking}})\simeq1000\sim1200$ fm using
  $r_{\text{breaking}}\simeq1.2\;-\;1.4$ fm from $2$-flavor lattice studies at
  $T=0$ and quark mass $m_\pi/m_\rho\simeq0.7$ \cite{Pennanen:2000yk}. In
  $2$-flavor lattice studies at $T=0$ and lower quark mass,
  $m_\pi/m_\rho\simeq0.4$, the string is expected to break at smaller (lower)
  distances (energies) \cite{Bernard:2001tz}. The thick line around $T_c$ is
  explained in Sec.~\ref{S}.}
\label{finf}
\end{figure}
Our results for $F_\infty(T)$ are summarized in Fig.~\ref{finf} as function of
$T/T_c$ and compared to $F_\infty(T)$ obtained in quenched and $3$-flavor QCD.
While $F_\infty(T)$ in quenched QCD exhibits a singularity at $T_c$ due to the
first order phase transition and is infinite below $T_c$, it is well-defined
and finite in full QCD at all temperatures due to string breaking below and
color screening above the transition. In this case $F_\infty(T)$ is steadily
decreasing with increasing temperatures in the whole temperature range analyzed
by us. A discussion of $F_\infty(T)$, in particular for $T\;\lsim\;T_c$, has
already been given \cite{Kaczmarek:2005ui,Kaczmarek:2005uv}. We may add here
that the slope of $F_\infty(T)$ as function of temperature indeed turns out to
be maximal in the vicinity of the transition. Although flavor and/or quark mass
dependences of the observable $F_\infty(T)$ can clearly be seen when comparing
the data for $2$- and $3$-flavor QCD, at temperatures about
$1.2T_c\;\lsim\;T\;\lsim\;2T_c$ no or only little differences between quenched
and full QCD could be identified. Flavor dependences are, however, to be
expected when reaching temperatures in the perturbative regime.

\subsection{The entropy}\label{S}
\begin{figure}[tbp]
  \epsfig{file=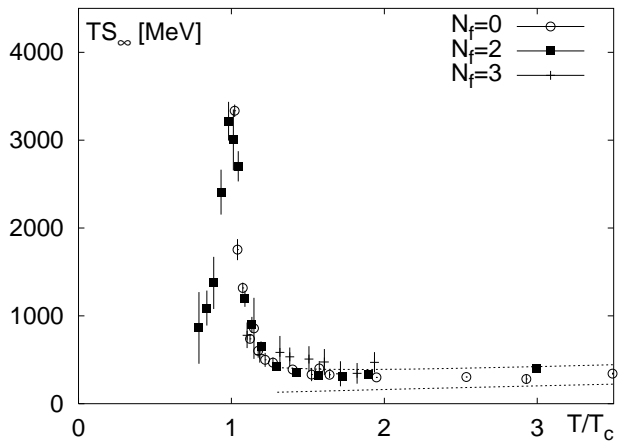,width=9.cm}
\caption{
  The contribution $TS_\infty(T)$ appearing in the free energy,
  $F_\infty(T)=U_\infty(T)-TS_\infty(T)$, calculated in $2$-flavor QCD as
  function of $T/T_c$. We compare our results from $2$-flavor QCD to the
  leading order part of Eq.~(\ref{entlimit}) (or Eq.~(\ref{ts})) using the
  $2$-loop formula for the coupling (\ref{2loop}) with
  $T_c/\Lambda_{\overline{\rm MS}}=0.77(15)$ \cite{Karsch:2000ps,Gockeler:2005rv} and
  scales $\mu=\pi,...,4\pi$ (dashed lines). We also show results from
  $3$-flavor QCD for $T\;\gsim\;1.1T_c$ \cite{Petreczky:2004pz} and quenched
  QCD \cite{Phd,Kaczmarek:2002mc,tobep}.  }
\label{entropy_all}
\end{figure}
The entropy contribution, $TS_\infty(T)$, obtained in $2$-flavor QCD is shown
in Fig.~\ref{entropy_all} as function of $T/T_c$ and is again compared to
results from quenched and $3$-flavor QCD. We indeed find $TS_\infty(T)\;\>\;0$
at all temperatures analyzed here. Moreover, the data for $TS_\infty(T)$ at low
temperatures also suggest a vanishing contribution in the zero temperature
limit, $T\to0$. Unfortunately we could not go to smaller temperatures to
clearly demonstrate this behavior. On the other hand, in the high temperature
phase, {\em i.e.} at $T\;\gsim\;2T_c$, we indeed find a tendency for an increase
of $TS_\infty(T)$ with temperature as expected from (\ref{ts}). Actually, this
small increase is consistent with the rise given in (\ref{entlimit}). To
demonstrate this we also compared $TS_\infty(T)$ in $2$-flavor QCD to
Eq.~(\ref{entlimit}) using the perturbative $2$-loop coupling, {\em i.e.}
\begin{eqnarray}
g_{2-loop}^{-2}(T)&=&2\beta_0\ln\left(\frac{\mu T}
{\Lambda_{\overline{\rm MS}}}\right)+\frac{\beta_1}{\beta_0}
\ln\left(2\ln\left(\frac{\mu T}{\Lambda_{\overline{\rm MS}}}\right)\right)\;,\nonumber\\
\label{2loop}
\end{eqnarray}  
with
\begin{eqnarray}
\beta_0&=&\frac{1}{16\pi^2}\left(11-\frac{2N_f}{3}\right)\;,\nonumber\\
\beta_1&=&\frac{1}{(16\pi^2)^2}\left(102-\frac{38N_f}{3}\right)\;,\nonumber
\end{eqnarray}
assuming vanishing quark masses. We used $T_c/\Lambda_{\overline{\rm MS}}=0.77(15)$
\cite{Karsch:2000ps,Gockeler:2005rv,Laine:2005ai} and the ambiguity in fixing
the scale in perturbation theory, $\mu=\pi,...,4\pi$. This estimate is shown
within the dashed lines and qualitatively agrees with the lattice data for
$T\;\gsim\;2T_c$. However, to clearly establish the perturbative increase of
$TS_\infty(T)$ with increasing temperature will require the analysis of
significantly higher temperatures. We also note that within the statistical
accuracy of the data we find for $T\;\gsim\;2T_c$ the tendency,
\begin{eqnarray}
S_\infty^{N_f=0}(T)\;\lsim\;S_\infty^{N_f=2}(T)\;\lsim\;S_\infty^{N_f=3}(T)\;.\label{tend}
\end{eqnarray}
It appears indeed quite reasonable that introducing additional flavor degrees
of freedom may enhance the finite temperature quark anti-quark entropy in the
deconfined phase. It is, however, quite difficult to separate clearly the
different effects from flavor and quark mass dependence in full QCD. In
particular, at temperatures below $T_c$ the tendency given in (\ref{tend}) may
change as can be seen from the temperature dependence of $F_\infty(T)$ shown in
Fig.~\ref{finf}.

In contrast to the small temperature dependence of $TS_\infty(T)$ at low and
high temperatures, $TS_\infty(T)$ shows qualitatively and quantitatively
significant differences at temperatures in the vicinity of the transition. In
fact, $TS_\infty(T)$ obtained in $2$-flavor QCD exhibits a sharp peak at $T_c$.
This behavior signals the high temperature phase transition/crossover in QCD.
As the peak is so sharp we may introduce $T_l$ ($T_u$) defined as the lower
(upper) temperature at which $S_\infty(T)$ approaches about half of the peak
value, {\em i.e.} $S_\infty(T_{l,u})\equiv S_\infty(T_c)/2$. We find for
$2$-flavor QCD $T_l$ about $0.89T_c$ and $T_u$ about $1.07T_c$ using
$S_\infty(T_c)\simeq16.5$. This temperature range is shown at the bottom of
Fig.~\ref{finf} as thick line. A similar behavior is also apparent in
$3$-flavor QCD. This indicates that the crossover from the low to the high
temperature phase in QCD takes place in a small temperature range around $T_c$.
We stress, however, that the temperature dependence of $F_\infty(T)$ also at
temperatures above $1.07T_c$ is still to a large extent dominated by
non-perturbative effects.

\subsection{The internal energy}
\begin{figure}[tbp]
  \epsfig{file=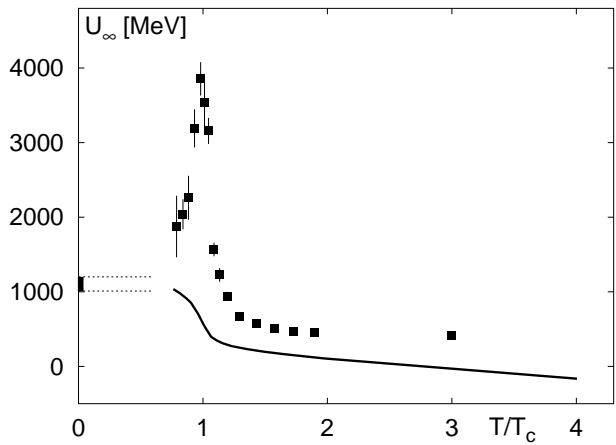,width=9.cm}
\caption{
  The internal energy $U_\infty(T)$ versus $T/T_c$ calculated in $2$-flavor
  QCD. The corresponding free energy, $F_\infty(T)$, calculated in $2$-flavor
  QCD is also shown as solid line. We again indicate in this figure the energy
  at which string breaking is expected to take place at $T=0$,
  $V(r_{\text{breaking}})\simeq 1000\;-\;1200$ MeV (dashed lines), using
  $r_{\text{breaking}}=1.2\;-\;1.4$ fm \cite{Pennanen:2000yk}.}
\label{string_breaking_paper}
\end{figure}
The internal energy, $U_\infty(T)$, in $2$-flavor QCD is shown in
Fig.~\ref{string_breaking_paper} as function of temperature and is compared to
the corresponding free energy, $F_\infty(T)$ (solid line), already shown in
Fig.~\ref{finf}. We indeed find $U_\infty(T)>F_\infty(T)$ at all temperatures
analyzed here. It can clearly be seen that the temperature dependence of
$F_\infty(T)$ and $U_\infty(T)$ is qualitatively and quantitatively different.
While the free energy steadily decreases with increasing temperatures the
internal energy exhibits a pronounced peak. Again this peak is sharply
localized at the (pseudo-) critical temperature. At the temperatures analyzed
by us the internal energy below $T_c$ is rapidly increasing with increasing
temperatures. Again we indicate by dotted lines the plateau value of the heavy
quark potential at zero temperature,
$V(r_{\text{breaking}})\simeq1000\;-\;1200$ MeV, using
$r_{\text{breaking}}\simeq1.2\;-\;1.4$ fm \cite{Pennanen:2000yk}. A comparison
of $U_\infty(T)$ with this value shows again that most of the temperature
dependence of $U_\infty(T)$ is sharply localized at temperatures in the
vicinity of the transition. A qualitatively similar behavior is also apparent
in $3$-flavor QCD \cite{Petreczky:2004pz}. Comparing the available data some
flavor or quark mass dependence, $U_\infty^{N_f=2}(T_c)\simeq 4000$ MeV $\gsim$
$U_\infty^{N_f=3}(T_c)\simeq3000$ MeV, can be observed and also the value
$U_\infty(T_c^+)$ in quenched QCD \cite{Phd} is of similar magnitude. As noted
in Sec.~\ref{exp}, however, the values for $U_\infty(T)$ may depend on the
relative normalization of $V(r)$ at $T=0$ used for renormalization.

At temperatures above $T_c$, $U_\infty(T)$ rapidly drops while at higher
temperatures, {\em i.e.} $T\;\gsim\;1.3T_c$, the temperature dependence of
$U_\infty(T)$ turns out to be much weaker than in the vicinity of phase
transition. However, contact with the perturbative relation, Eq.~(\ref{Upert}),
is not expected at those temperatures as $U_\infty(T)$ is still positive. In
fact, when approaching the perturbative high temperature regime also
$U_\infty(T)$ is expected to exhibit a change in sign and will slowly diverge
with respect to (\ref{Upert}).

\section{Finite temperature energies and potential models}\label{sec3}
\begin{figure}[tbp]
   \epsfig{file=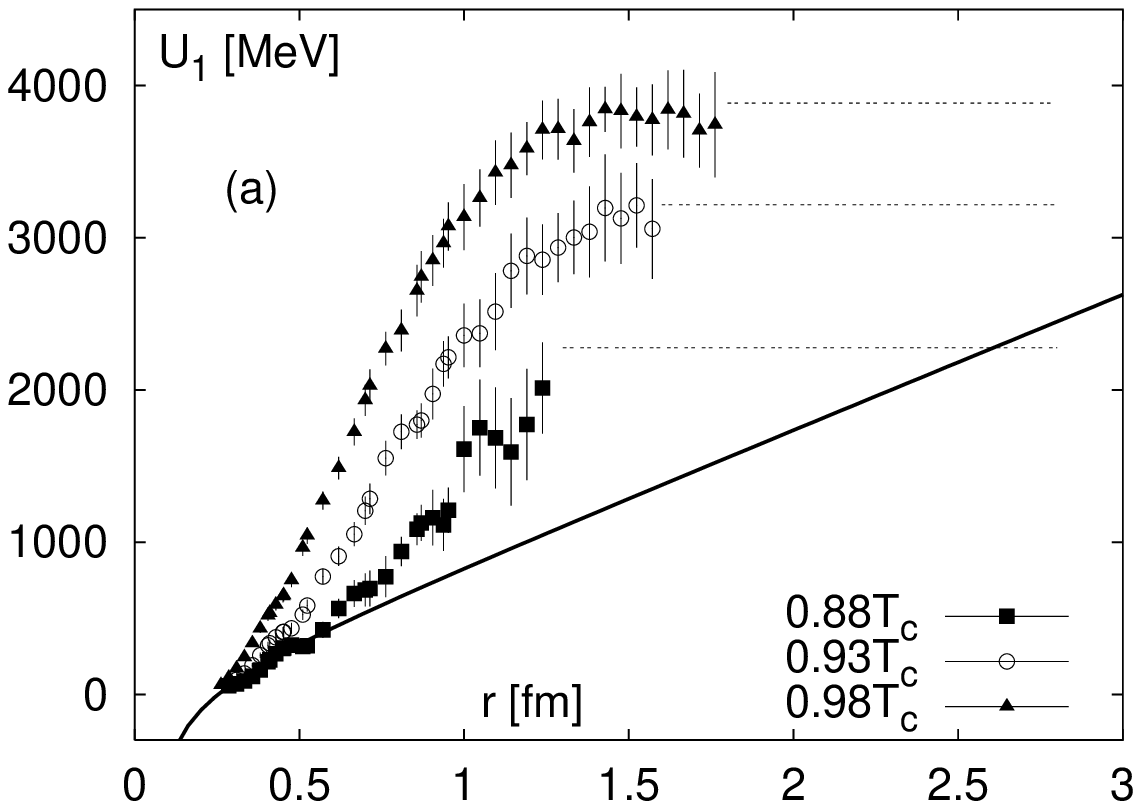,width=9.0cm}
   \epsfig{file=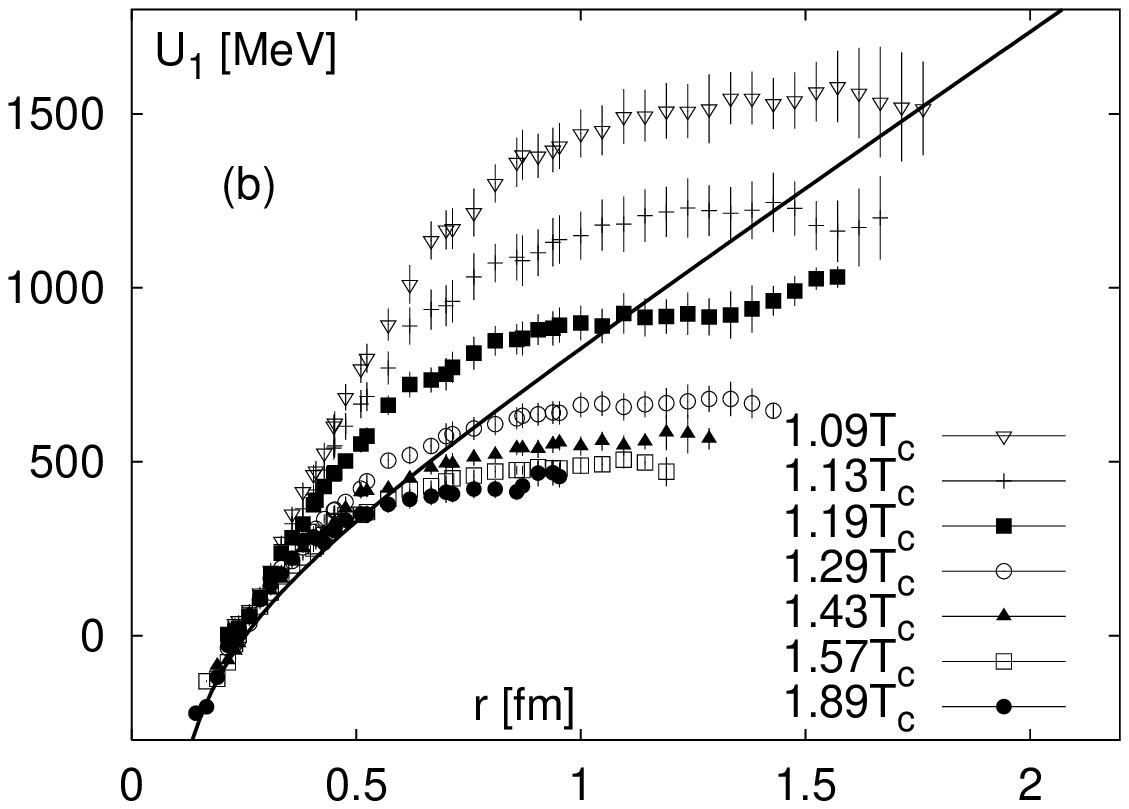,width=9.0cm}
\caption{The color singlet quark anti-quark internal energies, $U_1(r,T)$, at
   several temperatures below (a) and above (b) the phase transition obtained
   in $2$-flavor lattice QCD. In (a) we also show as horizontal lines the
   asymptotic values given in Tab.~\ref{tab1} which are approached at large
   distances and indicate the flattening of $U_1(r,T)$. The solid lines
   represent in each figure the $T=0$ heavy quark potential, $V(r)$
   \cite{Kaczmarek:2005ui,Kaczmarek:2005uw}.
}
\label{plc_en}
\end{figure}

The analysis of bound state problems has been quite successful in terms of
potential theory at $T=0$ \cite{Eichten:1978tg,Eichten:1979ms,Jacobs:1986gv}.
For the discussion of quarkonium suppression patterns at finite temperature one
also often resorts to potential models
\cite{Digal:2001iu,Digal:2001ue,Wong:2004kn,Shuryak:2003ty,Park:2005nv}. Of
course, the strong interaction remains unaffected by temperature and the
modeling of thermal modifications of heavy quark bound states requires the
definition of an effective potential, $V_{\text{eff}}(r,T)$
\cite{Karsch:2005ex}, which can be given only phenomenologically, for instance,
by using the modifications of the free and internal energies. It is thus
important to understand the binding properties of the different finite
temperature energies. For this purpose we also calculated the quark anti-quark
internal energies for the temperatures given in Tab.~\ref{tab1} at several
finite distances. Parts of our results for $U_1(r,T)$ are summarized in
Fig.~\ref{plc_en} at temperatures below (a) and above (b) the transition.
Similar results have been obtained for internal energies in the averaged and
octet channels.

\begin{figure}[bp]
  \epsfig{file=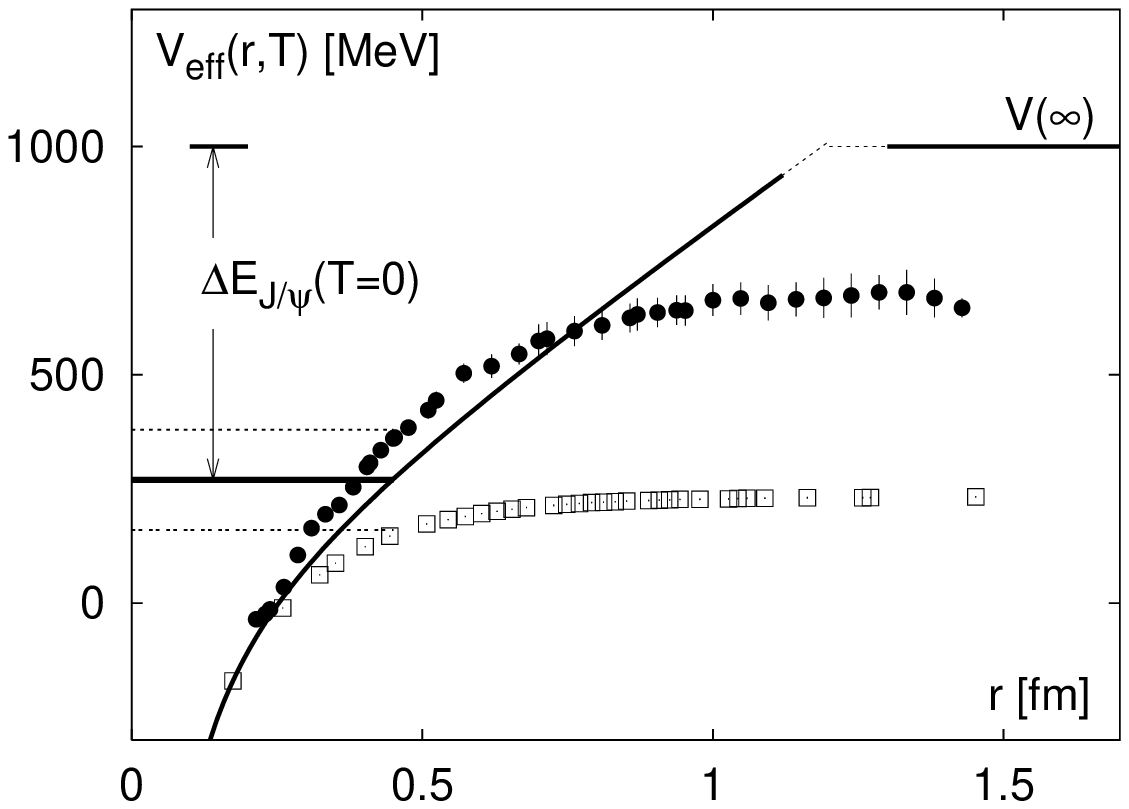,width=9.0cm}
\caption{
  Different effective potentials in the color singlet channel,
  $V_{\text{eff}}(r,T)\equiv F_1(r,T)$ (open symbols) and
  $V_{\text{eff}}(r,T)\equiv U_1(r,T)$ (filled symbols), at fixed
  $T\simeq1.3T_c$ as function of distance and the heavy quark potential,
  $V(r)$. We also compare $V_{\text{eff}}(r,T)$ to the $J/\psi$ energy level at
  $T=0$, $E_{J/\psi}(T=0)\equiv V(r=r_{J/\psi})\simeq270$ MeV (horizontal solid
  line). The horizontal dashed lines correspond to the $J/\psi$ energy levels
  defined on $E_{J/\psi}(F_1)\equiv F_1(r_J/\psi,T)$ (lower dashed line) and
  $E_{J/\psi}(U_1)\equiv U_1(r_{J/\psi},T)$ (upper dashed line).
  $r_{J/\psi}\simeq 0.45$ fm is fixed through the mean squared charge radius
  expected at $T=0$ given in Tab.~\ref{tab}.  }
\label{EB}
\end{figure}
To gain some insight into the consequences these energies have for quarkonium
dissociation we compare the asymptotic ($r\to\infty$) energies with the
potential energies at a distance corresponding to the size of some quarkonium
states,
\begin{eqnarray}
\Delta E_i(T=0)\equiv V(\infty)-V(r_i)\;,
\end{eqnarray}
where the radii $r_i$ ($i=J/\psi,\;\chi_c,...$) are listed in the first row of
Tab.~\ref{tab}. At zero temperature $V(\infty)$ is taken to be twice the energy
needed to create the lowest heavy-light meson. For our purpose we consider
$V(\infty)\equiv V(r_{\text{breaking}})\simeq1000$ MeV where
$r_{\text{breaking}}$ is the distance at which the string is expected to break
at zero temperature, $r_{\text{breaking}}\;\gsim\; 1.2$ fm
\cite{Pennanen:2000yk}. This energy is shown in Fig.~\ref{EB} as horizontal
line. The resulting energy for $J/\psi$ is also shown. The energies for some
charmonium and bottomonium states are summarized in Tab.~\ref{tab} and compared
to the mass difference obtained from $2M_{D,B}-m_{i}$ where $M_{D,B}$ denotes
the $D$- and $B$-meson masses and $m_i$ the masses of the different quarkonium
states \cite{Eidelman:2004wy}. Of course, the wave functions for the different
quarkonium states will also reach out to larger distances \cite{Jacobs:1986gv}
and thus our estimate for the different energy levels $E_i(T=0)$ can only be
taken as indicative for the relevant energies. Potential model analysis, using
for instance the Schr\"odinger equation, will do better in this respect.
\begin{figure}[tbp]
  \epsfig{file=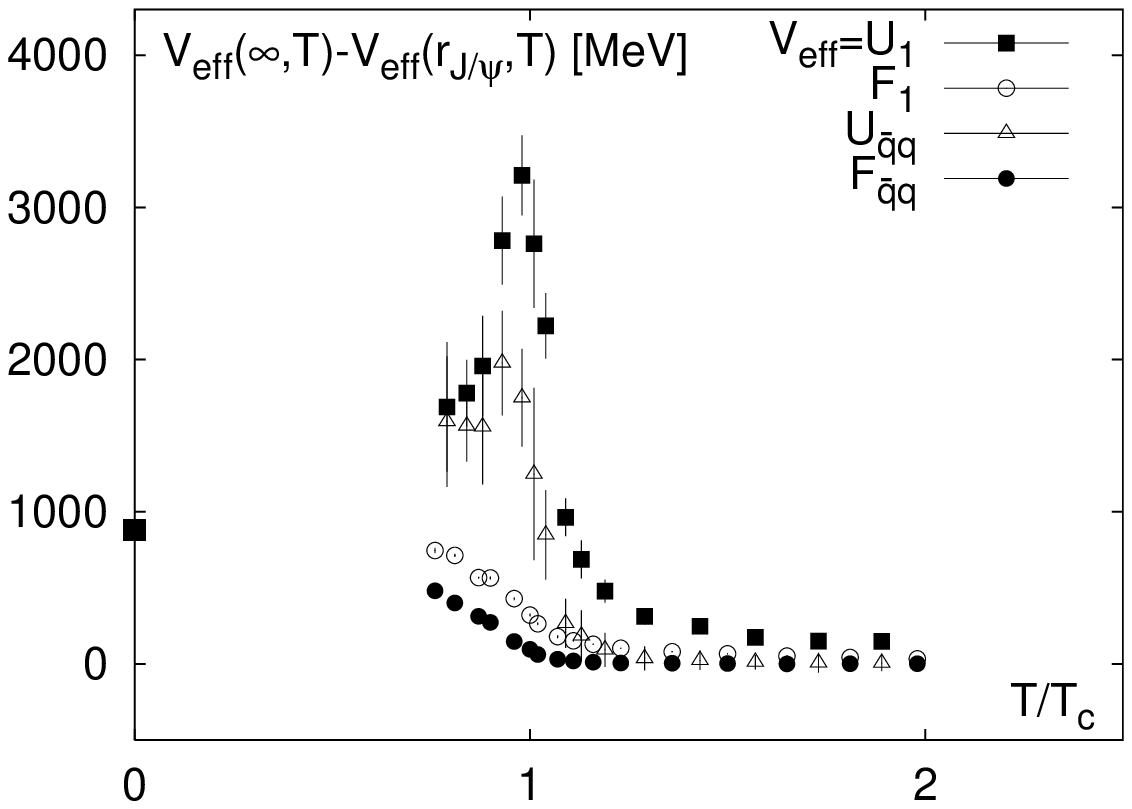,width=9.0cm}
\caption{
  The energy difference $V_{\text{eff}}(\infty,T)-V_{\text{eff}}(r_{J/\psi},T)$
  as function of temperature using the different energies as effective
  potential, $V_{\text{eff}}(r,T)\equiv U_1(r,T),\;F_1(r,T),\;U_{\bar q
    q}(r,T),\;F_{\bar q q}(r,T)$.  }
\label{EBin}
\end{figure}

\begin{table*}[htbp]
\centering
\setlength{\tabcolsep}{0.7pc}
\begin{tabular}{|l|lll|lll|}
\hline
state & $J/\psi$ & $\chi_c$ & $\psi\prime$ & $\Upsilon$ & $\chi_b$ & $\Upsilon\prime$ \\
\hline
\hline
$r_i$ [fm] & 0.45 & 0.70 & 0.87 &0.23 & 0.41 & 0.51 \\
$2M_{D,B}-m_i$ [MeV] &631 & 313&42 & 1098&698 &535 \\
$\Delta E_i(T=0)$ [MeV] & 830 & 560& 400 &1140 &870 &750\\
\hline
\hline
$\Delta E_i(T_c)$ [MeV] [using $F_1$] & 321 & 124 & 62 & 665 & 334 & 265\\
$\Delta E_i(T_c)$ [MeV] [using $U_1$] &3036 &1618 & 1010&3721 &3085 & 2615  \\
$\Delta E_i(T_c)$ [MeV] [using $F_{\bar q q}$] & 97& 27 & 12 & 607 & 87 &  55\\
$\Delta E_i(T_c)$ [MeV] [using $U_{\bar q q}$] &1355 &747 &504 & 3732& 1392&1136 \\
\hline
\hline
$T^{\text{PM}}_{\text{dis}}/T_c$ [potential model using $F_1$] &1.1 &0.74 &0.1 - 0.2 & 2.31 & 1.13 & 1.1\\
$T^{\text{PM}}_{\text{dis}}/T_c$ [potential model using $U_1$] &$\sim2$ &$\sim1.1$ &$\sim1.1$ &$\sim4.5$ &$\sim2$ &$\sim2$ \\
\hline
\hline
$T^{\text{SF}}_{\text{dis}}$ from lattice spectral functions &$1.5\;-\;3$ &$\lsim\;1.1$ & & & & \\
\hline
\end{tabular}
\caption{A summary of the different estimates for the relevant temperatures,
  $T_{\text{dis}}$, at which quarkonium dissociation is expected to become
  important. In the first row we give the mean squared charge radii, $r_i$, and
  the energies, $\Delta E_i(T=0)\equiv V(\infty)-V(r_i)$ at zero temperature
  with $V(\infty)\simeq1000$ MeV. We also list the mass gap $2M_{D,B}-m_i$
  using $D$- and $B$-meson masses $M_D=1864$ MeV and $M_B=5279$ MeV,
  respectively \cite{Eidelman:2004wy}. The second row summarizes results for
  the break up energies at $T_c$ estimated from $\Delta E_i(T_c)\equiv
  V_{\text{eff}}(r_i,T_c)-V_{\text{eff}}(\infty,T_c)$ using different
  definitions for $V_{\text{eff}}(r,T)$. The errors on these values are
  typically about $20\%$. Note here that the thermal energy $E_{th}(T)=3T$ is
  about 600 MeV at $T_c$. The $3^{\text{rd}}$ and $4^{\text{th}}$ row contain
  different values for the relevant dissociation temperatures, {\em i.e.}
  estimates for the onset of temperature effects in quarkonium states,
  $T_{\text{onset}}$ \cite{Kaczmarek:2005ui,Karsch:2005ex}, the predicted
  dissociation temperatures from potential model calculations
  ($T^{PM}_{\text{dis}}$) \cite{Digal:2001iu,Digal:2001ue,Wong:2004kn}, and
  lattice studies of charmonium spectral functions,
  $T^{\text{SF}}_{\text{dis}}$ \cite{Asakawa:2003re,Datta:2003ww}.}
\smallskip
\label{tab}
\end{table*}

Similarly we can estimate the temperature dependence of the energy levels for
the different quarkonium states from $E_i(T)\equiv V_{\text{eff}}(r_i,T)$.
Again these energy levels will only characterize the relevant energies and the
sizes of these states may also become temperature dependent. At finite
temperature, however, the values for these levels are expected to depend
crucially also on the specific modeling of the effective potential,
$V_{\text{eff}}(r,T)$. This is obvious from the different energy levels for the
$J/\psi$ shown in Fig.~\ref{EB} which we obtained by using as
$V_{\text{eff}}(r,T)$ the singlet free energy (lower dashed line) and singlet
internal energy (upper dashed line). Due to the steeper rise of the internal
energy compared to free energy $E_{J/\psi}(T)$ is enhanced compared to the
energy level obtained from the internal energy. It is interesting to note here
that $E_{J/\psi}(T=1.3T_c)$ deduced from the internal energy is even larger
than at zero temperature while in terms of the free energy it is smaller than
$E_{J/\psi}(T=0)$. For the characterization of the relevant energies needed to
dissociate the bound state we again consider $\Delta E_i(T)\equiv
V_{\text{eff}}(\infty,T)-V_{\text{eff}}(r_i,T)$, which depends on the
definition of $V_{\text{eff}}(r,T)$. This is evident from Fig.~\ref{EBin} where
we show the temperature dependence of $\Delta E_{J/\psi}(T)$ estimated from
$V_{\text{eff}}(r,T)\equiv U_1(r,T)$ (filled circles) and
$V_{\text{eff}}(r,T)\equiv F_1(r,T)$ (open circles). While $\Delta
E_{J/\psi}(T)$ is continuously decreasing with increasing temperatures when
using $F_1(r,T)$ as effective potential, its binding pattern appears quite
different from what one obtains by using $V_{\text{eff}}(r,T)\equiv U_1(r,T)$.
Actually, in the latter case the binding pattern exhibits a maximum in the
vicinity of the transition while it rapidly drops above $T_c$. Similar results
have been obtained also for other charmonium and bottomonium states and are
summarized in Tab.~\ref{tab}.

Of course, the model dependences of the dissociation energies at finite
temperature also affect the analysis of suppression patterns and corresponding
dissociation temperatures \cite{Digal:2001iu,Digal:2001ue,Wong:2004kn}.
Actually, using $V_{\text{eff}}(r,T)\equiv U_1(r,T)$ suggests that suppression
of $J/\psi$ may occur only at temperatures close but above the transition while
from $V_{\text{eff}}(r,T)\equiv F_1(r,T)$ one finds that $J/\psi$ dissolves
already at temperatures below the crossover. Similar model dependences enter
also the analysis of excited quarkonium states and the corresponding estimates
for the dissociation temperatures are summarized in Tab.~\ref{tab} using four
different definitions for $V_{\text{eff}}(r,T)$, {\em i.e.} we used the singlet
free and internal energies ($F_1(r,T)$, $U_1(r,T)$) as well as the finite
temperature energies in the color averaged channel ($F_{\bar q q}(r,T)$,
$U_{\bar q q}(r,T)$).

\section{Summary and conclusions}\label{seccon}
Following \cite{Kaczmarek:2002mc,Zantow:2003ui} we introduced and analyzed the
change in internal energy and entropy due to the presence of a static quark
anti-quark pair in a QCD heat bath. Both observables are introduced as
intensive observables as appropriate derivatives of the renormalized free
energy. Similar to the singlet quark anti-quark free energies
\cite{Kaczmarek:2002mc,Zantow:2003uh,Zantow:2001yf,Kaczmarek:2005ui} also the
singlet internal energies become temperature independent in the limit of small
distances and are controlled by the zero temperature running coupling.

We analyzed qualitative and quantitative differences that appear when changing
from free energies to internal energies as observable that defines an effective
potential that can be used in model calculations. We discussed the important
role of the entropy contribution at finite temperature. At short distances, at
intermediate and at large distances the entropy contribution is non-zero and
shows non-trivial $r$-dependences. We find positive entropy contributions,
$S_1(r,T)\;\gsim\;0$, and thus $U_1(r,T)\;\gsim\;F_1(r,T)$. Similar to the free
energies, also the large distance properties of the internal energies and
entropies are controlled by string breaking below and color screening above
deconfinement and both approach temperature dependent constant values, which
define $U_\infty(T)$ and $S_\infty(T)$, at asymptotic large distances.
Actually, the difference between free and internal energies at high
temperatures, $TS_\infty(T)\simeq 4m_D(T)\alpha(T)/3$, is supposed to increase
with increasing temperatures. In particular, $U_\infty(T)$ and $S_\infty(T)$,
are, similar to $F_\infty(T)$ \cite{Kaczmarek:2002mc}, again introduced as
manifest gauge invariant observables and clearly signal the QCD plasma
transition. In fact, while the plateau values which are approached by the free
energies, $F_\infty(T)$, are rapidly decreasing in the vicinity of the
transition \cite{Kaczmarek:2005ui,Kaczmarek:2005uv}, the values approached by
the internal energies, $U_\infty(T)$, and entropies, $S_\infty(T)$, show both a
sharp peak at the (pseudo-) critical temperature. Similar results are also
obtained in quenched and $3$-flavor QCD
\cite{Kaczmarek:2002mc,Zantow:2003ui,Phd,Petreczky:2004pz}. However,
qualitative differences become quite transparent in the vicinity and below the
transition when comparing these observables obtained in quenched and full QCD.
In quenched QCD the first order phase transition is related to singularities in
thermodynamic observables which can indeed be seen after renormalization in the
temperature dependence of the finite temperature energies, entropies and the
renormalized Polyakov loop (see also
\cite{Kaczmarek:2002mc,Kaczmarek:2005ui,tobep}). In contrast to quenched QCD,
in full QCD the phase change is a crossover and we consequently do not see any
singularities in the finite temperature energies and entropies nor in the
temperature dependence of the renormalized Polyakov loop
\cite{Kaczmarek:2005ui}.

We also investigated the temperature dependence of quarkonium binding in the
vicinity of the transition. For this purpose we used the finite temperature
energies to define appropriate effective potentials, $V_{\text{eff}}(r,T)$, as
one would do in potential models. As effective potentials we used the singlet
and averaged free and internal energies, {\em i.e.} we defined
$V_{\text{eff}}(r,T)$ through $F_1(r,T)$, $F_{\bar q q}(r,T)$, $U_1(r,T)$ and
$U_{\bar q q}(r,T)$, and estimated the binding energies at temperatures below
and above the transition. In all cases the binding energies of the quarkonium
states indeed become weaker with increasing temperatures above the transition
and this may lead to dissociation of parts of these states at temperatures
close but above $T_c$. Our analysis, however, shows strong dependencies of the
binding energies and dissociation temperatures on the specific modeling of
$V_{\text{eff}}(r,T)$. To some extent these model dependencies enter from quite
general grounds when exchanging the definition of $V_{\text{eff}}(r,T)$ from
free energies into internal energies. When using free energies the binding
energies continuously and rapidly decrease when crossing the transition and
most of the quarkonium bound states may thus indeed dissolve at temperatures in
the vicinity and below the transition \cite{Digal:2001iu,Digal:2001ue}. The
temperature dependence of the binding energies deduced from internal energies,
however, turns out to be more complicated in the vicinity of the transition. An
effective potential defined through quark anti-quark internal energies suggests
increasing binding energies below the transition which exhibit a peak at $T_c$.
This may imply that all quarkonium states analyzed here are still bound at the
transition. The binding energies of the different states rapidly decrease above
$T_c$ leading again to quarkonium dissociation, however, the temperatures which
are relevant for dissociation are shifted to larger temperatures than those
deduced from free energies. The recent potential model calculations
\cite{Shuryak:2004tx,Wong:2004kn} using the properties of $U_1(r,T)$ at
temperatures above the transition as well as the lattice analysis of quarkonium
bound states in quenched QCD \cite{Asakawa:2003re,Datta:2003ww} support our
findings.

\begin{acknowledgments}
  We thank the Bielefeld-Swansea collaboration for providing us their
  configurations with special thanks to S. Ejiri. We would like to thank E.
  Laermann, F. Karsch and H. Satz for many fruitful discussions. F.Z. thanks P.
  Petreczky for his continuous support. We thank K. Petrov and P. Petreczky for
  sending us the data of Ref.~\cite{Petreczky:2004pz}. This work has partly
  been supported by DFG under grant FOR 339/2-1 and by BMBF under grant
  No.06BI102 and partly by contract DE-AC02-98CH10886 with the U.S. Department
  of Energy. At an early stage of this work F.Z. has been supported through a
  stipend of the DFG funded graduate school GRK881.
\end{acknowledgments}

\bibliographystyle{h-physrev3} \bibliography{paper}

\end{document}